# Hierarchical self-organization of tectonic plates

G. Morra, M. Seton, R.D. Müller

Earthbyte Group, School of Geosciences, University of Sydney

The Earth's surface is subdivided into eight large tectonic plates and many smaller ones. We reconstruct the plate tessellation history and demonstrate that both large and small plates display two distinct hierarchical patterns, described by different power-law size-relationships. While small plates display little organisational change through time, the structure of the large plates oscillate between minimum and maximum hierarchical tessellations. The organization of large plates rapidly changes from a weak hierarchy at 120–100 million years ago (Ma) towards a strong hierarchy, which peaked at 65–50 Ma, subsequently relaxing back towards a minimum hierarchical structure. We suggest that this fluctuation reflects an alternation between top and bottom driven plate tectonics, revealing a previously undiscovered tectonic cyclicity at a timescale of 100 million years.

Plate tectonic cycles occur at several timescales, with seismic cycles in the order of hundreds to thousands of years, and cycles of several hundreds of million years (Myrs) for the formation and break-up of supercontinents such as Pangea and Rodinia (Wilson cycle). Regional plate tectonics is driven by plate subduction, in which slabs require 5–15 Myrs to reach the boundary between the Earth's upper and lower mantle; a time period also required to open back-arc basins. The intermediate timescale, between a few tens and a few hundred million years, is not yet well understood. In this timeframe, apparent polar wander studies and plate reconstructions have identified plate reorganisations at intervals of about 50 Myrs (*1*), and sea-level studies have identified cycles at the timescale of 10–100 Myrs (*2*). Whether tectonic cycles at the 100 Myrs scale exist and how they are related to geodynamic processes is still unclear. Here we use a novel approach to unravel how the time-dependent structure of plate tessellations is the surface manifestation of two different modes of mantle convection.

The Pacific Plate is the largest tectonic plate on Earth, followed by Africa, which is the antipode of the Pacific and mostly surrounded by mid-ocean ridges (Fig. 1). Combined with another six major plates (Eurasia, Australia, Antarctica, North and South America, and Nazca) this set of large plates dominates the surface of the Earth (*5*). The remaining plates are one or more orders of magnitude smaller and their size has different statistical properties (*6*), probably reflecting a fragmentation mechanism (*7*). The main mechanism influencing the size of the largest plates is controversial: size could be driven from the underlying convecting mantle (*8*), or from the same fragmentation process that forms the smaller plates (*7*).

Complex processes such as fragmentation are often revealed by geometrical self-similarity, where a reiteration of patterns occurs over a wide range of scales. Such a multiscale property is conventionally described by a power-law, a function of the form $S(x) \propto x^{-\alpha}$, where $S$ is the observable at the scale *x* and the exponent $\alpha$ is a scale-independent coefficient. The simplest way to measure $\alpha$ is to visualise it by a straight line in a log-log plot and quantify its slope, as in Figure 1 where two distinct slopes relate *plate size* with the *hierarchy* determined by the size itself: *$\alpha_{SP}$=4.5* for the small plates and *$\alpha_{LP}$=0.3* for the large plates. If such clearly distinct power-laws exist for large and small plates, they must originate from different dynamical processes (*9*). The established test of the existence of a power law relationship requires hundreds of data points spanning several orders of magnitude. The limitation of the conventionally used present-day data set is overcome by extending the data back in time and testing a small set of available plate boundary reconstructions for the past 140 Myrs in one million year intervals (details in the supplementary section).

The propagation of errors in plate boundary reconstructions back in time has a greater effect on the assessment of the size of small plates compared to larger ones. We therefore calculate $\alpha_{SP}$ only for the past 50 Myrs, when at least ten small plate polygons have been reconstructed, comparing 6 plate-size subsets to test statistical consistency (Fig. 2b). All 300 fits are found in the interval $\alpha_{SP}$=[3–5], which is in agreement with a fragmentation hypothesis ($\alpha$=4) (*6, 7*).

Large plates represent a compact set of 7 to 9 plates. It is possible to extend the calculation of $\alpha_{LP}$ back to 140 Ma because the relative error in plate size determination is proportionally inverse to size, and large plates cover a significantly greater area than small plates. We find that $\alpha_{LP}$ varies in the interval [0–1] and that the residuals of the fit diminish towards the present, when plate reconstructions are more reliable, thus supporting the detection of a power-law fit (see supplement). The clear separation between $\alpha_{LP}$ and $\alpha_{SP}$ provides strong evidence that the full range of tectonic plates is better described by two power-law distributions, where $\alpha_{LP}$ and $\alpha_{SP}$ mirror different plate evolution dynamics (Fig 2c).

There is a precise relationship between the magnitude of $\alpha_{LP}$ and the respective Earth surface tessellation (Figure 2a). When $\alpha_{LP}$ is at its minimum (almost zero, ~110 Ma) there is virtually no difference between plate sizes. Conversely, when $\alpha_{LP}$ reaches its maximum (almost one, ~60 Ma), corresponding to a large plate being surrounded by smaller plates, the power-law defines the relationship between a plate and the next smaller plate. We call the first configuration *weakly hierarchical* and the second *strongly hierarchical*. Such patterns mark a collective adaptation of the plates in a self-organised state because plates must fragment or merge to maintain the power law fit.

We propose that the evolution of $\alpha_{LP}$ implies the existence of a previously unmapped tectonic cycle at the time scale of 100 Myrs, characterised by a detailed structure resembling an impulse. This is illustrated in Figure 3 where $\alpha_{LP}$ departs from a weak hierarchical state at 110 Ma, rising in magnitude until 60 Ma, where it begins to slowly decline, continuing until presently.

There is striking geological evidence that this process represents a new plate tectonic cycle: plate break-up and plate amalgamation correlate with excitation and relaxation phases of $\alpha_{LP}$. During the excitation stage, two main events accommodated the growth $\alpha_{LP}$: the break-up of Australia and East and West Antarctica (83 Ma), and the inception of Kula-Pacific and Kula-Farallon spreading (79 Ma). At the end of this phase the tectonic system was completely reorganised: a record number of nine plate aligned their sizes to fit the power-law. Several events accommodated the relaxation phase: the Farallon-Phoenix ridge stopped spreading at 67 Ma, Indian and Australian plates merged at 43 Ma, and East and West Antarctica rejoined at 27 Ma. During the last 30 Myrs, the tectonic system has continued to relax while $\alpha_{LP}$ has decreased from 0.5 to 0.4.

Our analysis further reveals commonalities of causes between the two main plate reorganisations in the last 140 Myrs (at ~100 Ma and ~50 Ma), and identifies two turning points in the hierarchical statistics, namely at ~120–100 Ma and ~65–50 Ma. Such plate reorganisations have been global or have involved at least the Pacific and the neighbouring plates. The cause of the reorganization at ~50 Ma, clearly visible in bends of Pacific hotspot tracks (*10*), is likely due to the subduction of the ridge between the Izanagi and Pacific plates (*11*). At the same time this event marks the beginning of subduction of the Pacific Plate, which started to reduce size, toward a new plate hierarchy. This combination illustrates a mechanism through which the evolution of plate hierarchy can provide a signature of plate reorganisations.

The ~100 Ma reorganisation has been less well explored. Clear bends dated between 120–80 Ma have been observed in several hot spot tracks in the western Pacific (*10*) and in fracture zones in the eastern Indian ocean (*12*). It is possible that another ridge subduction, that of the Phoenix-Pacific plate boundary, caused this event (*13*), however, the replacement of the Izanagi Plate with the Pacific Plate as the leading plate in the hierarchy must have generated a new set of plate boundary forces. This would have propelled the tectonic system from a weak to a strong *hierarchical* state, and reorganised the Indian and Pacific plates into a new kinematic regime.

The evolution of an alternating weak and strong hierarchical system has a fluid-dynamic analogy that puts the top vs. bottom-driven plate tectonics controversy in a new light. Rayleigh-Benard, or *Benard,* convection is defined by the overturning of fluid in box heated from its interior or from the bottom, and represents a classical analogy of a convecting mantle. However, in this model the strength of the surface plates does not play a role; this notion has another analogy in fluid-dynamics, which is surface tension. Laboratory and numerical studies of convection have shown that surface tension can control size and shape of convective cells (*14*). In this case, convection is called Benard-Marangoni or simply *Marangoni*. While *Benard* surface cells have homogeneous sizes, *Marangoni* convection can exhibit non-homogeneous cell shapes (*15*). Two non-dimensional numbers control the role of the surface tension and heat forcing: the Marangoni and Rayleigh numbers ($M$ and $R$) (*16*), representing the balance of viscous dissipation with surface tensional forces ($M$) and buoyancy ($R$), respectively. While the separate assessment of $R$ and $M$ is very uncertain, their ratio $R/M$ is easier to determine and indicates the dominant forcing. The critical numbers for the onset of convection are $R_{oc}=680$ and $M_{oc}=81$, therefore the critical ratio is about $R_{oc}/M_{oc}=10$: if $R/M >>10$, buoyancy dominates (Benard convection); if $R/M<<10$, surface tension prevails (Marangoni convection).

For the Earth the ratio $R/M=g\alpha\rho d^2/\gamma$ is calculated from gravity ($g\sim 10$ $m/s^2$), thermal expansion ($\alpha\sim 4\cdot 10^{-5}$ $K^{-1}$), density ($\rho\sim 3000$ $Kg/m^3$), thickness of the convective domain ($d=600–2800$ $km$), and from the temperature derivative of the surface tension ($\gamma$). $\gamma$ can be derived by its analogy with the stiffness of the lithospheric core. This 10–30 km thick layer of the oceanic plate controls plate strength (*17*), the radius of curvature during subduction (*18*), and the plate's predisposition to fragment (*19*). At the first order $\gamma$ is the ratio between the plate stresses, between ridge push and the slab pull ($\sigma\sim 3\cdot 10^{12}–5\cdot 10^{13}$ $Pa\cdot m$) and the temperature variation in the lithosphere ($\Delta T\sim 500–1000$ K), which results in the range $\gamma=3\cdot 10^9–3\cdot 10^{11}$ Pa·m/K. The range of the resulting ratio $R/M=[4–3000]$ includes values close to and far away from the critical $R/M=10$. This means that the plate tectonic system also fluctuates between being close to and far away from the critical threshold, suggesting that the transition that we observe between weak *(~110 Ma)* and strong *(~60 Ma)* hierarchies mirrors the fluctuation between Benard and Marangoni convection styles, equivalent to bottom- and top-driven plate tectonics (Figure 3).

The implications of this previously unmapped, yet fundamental, global tectonic cyclicity span several aspects of geodynamics. It introduces a cyclic time-variation in the reconstruction of the thermal history of the Earth and offers an alternative explanation for the second-order sea-level cycles (*20*). Fragmentation models can be employed to push plate reconstruction further back in time, testing the statistics of completely subducted ocean basin sizes. Open questions related to Earth history include how the oscillations between Benard and Marangoni convection have affected global intraplate stresses and therefore paleoseismicity and tectonics (*21*), whether there is a common cause of the Pacific-Africa plate size polarity and the corresponding deep mantle low-velocity anomalies (*22*) and if there is a relationship between the minimum plate hierarchy and the cessation of magnetic field reversals during the Cretaceous superchron (*23*).

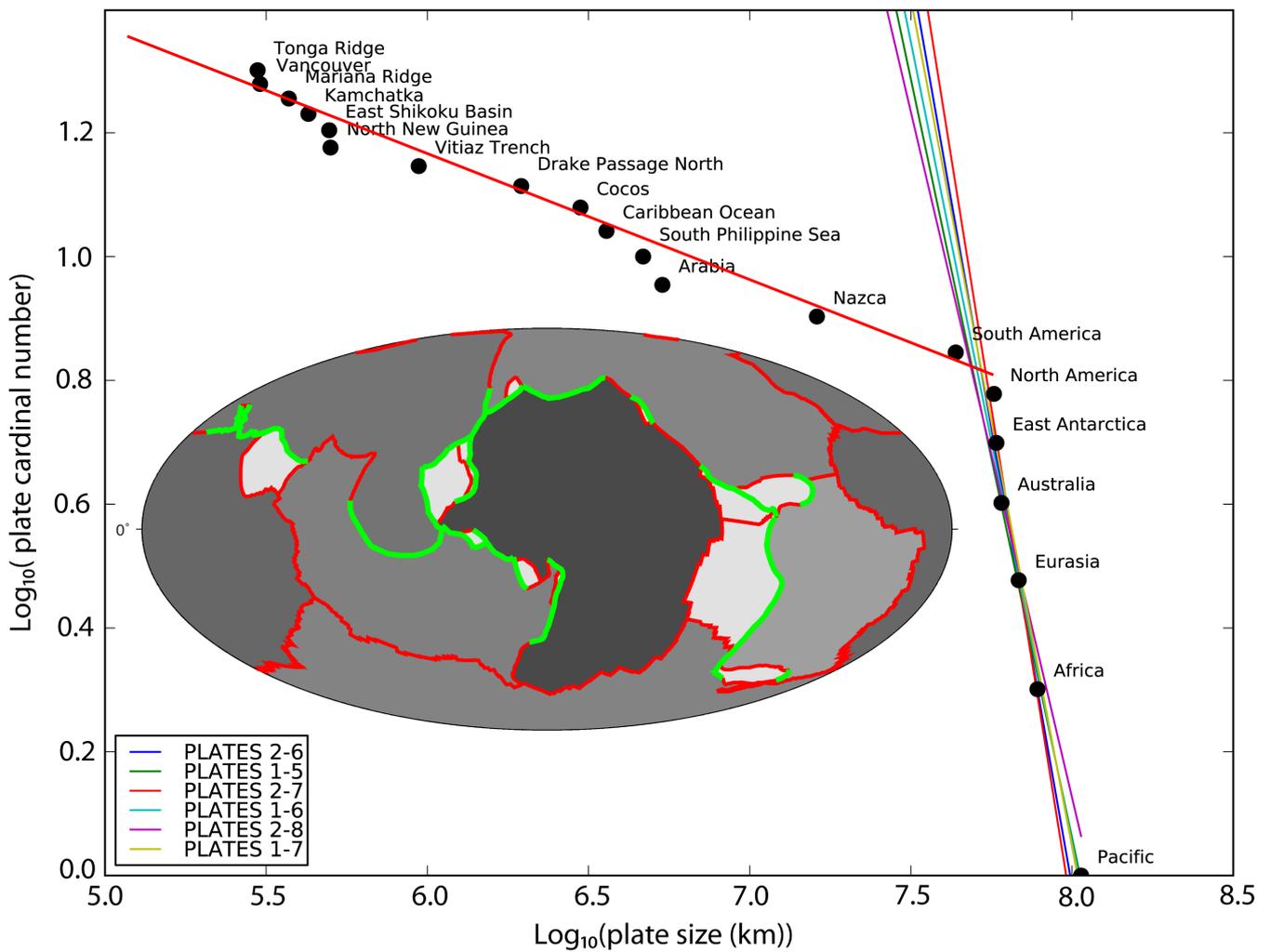

Figure 1. The planetary map represents the Earth's present tessellation. Plate colours are in Gray scale, where the BW gradation is proportional to the logarithm of plate size. The map illustrates the polarity of the tectonic system in which the two greatest plates, the African and Pacific, are at the antipodes of one another. The x-y plot represents the logarithm of the plate size (X axis) vs. logarithm of complementary cumulative number (Y axis) of the 22 plates shown in the global map. The coloured lines are based on 5 fits of overlapping subsets of plates. Two regimes are identifiable: (i) below log(Size)~7 the small plates follow a power-law with an exponent α~3–5, (ii) the largest 6–7 plates follow a steeper slope with α=0.3.

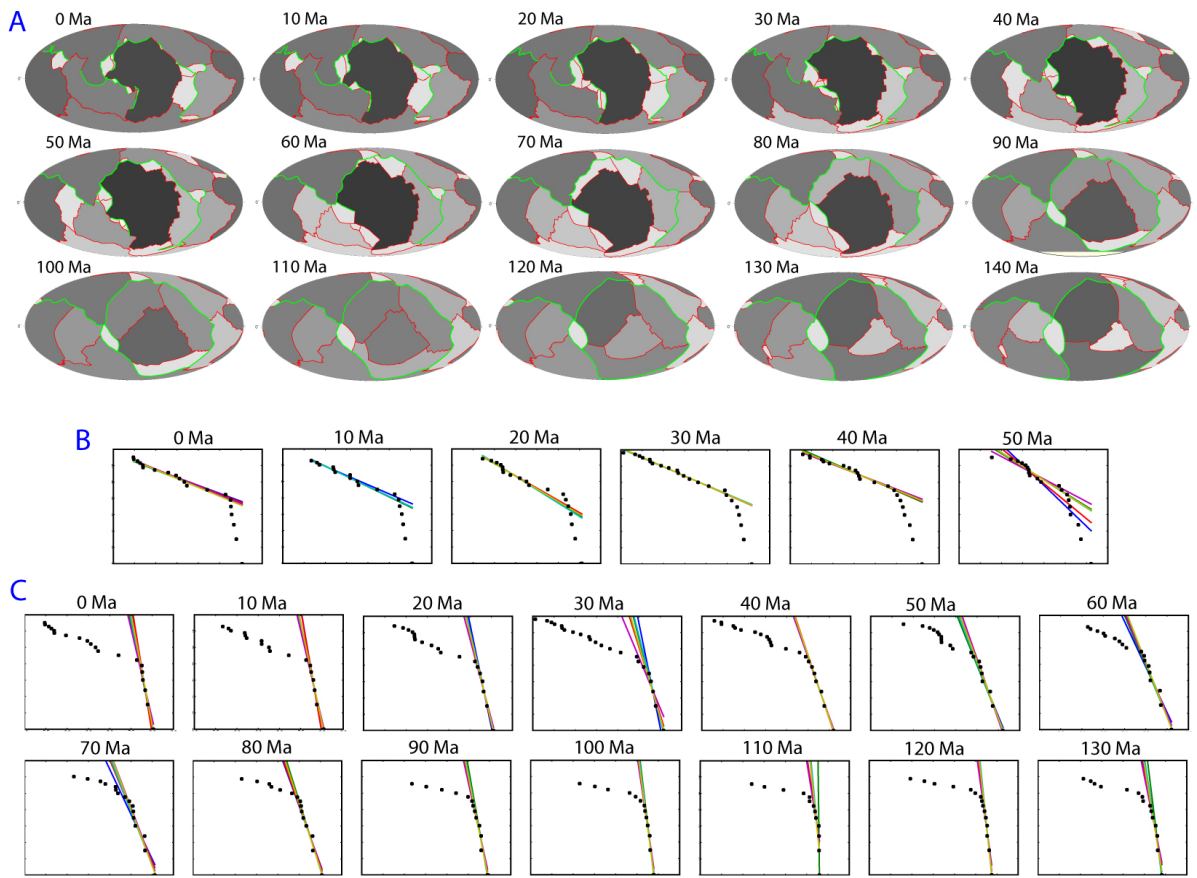

A: Earth Surface tessellation for the last 140 Myrs: darker grey indicates larger plate size. Strongly hierarchical tessellations have a large-dark Pacific (or Izanagi) Plate surrounded by smaller grey plates. Weakly hierarchical and uniform tessellations are uniformly grey. Green boundary lines are convergent margins while red boundaries are spreading. B: complementary cumulative number vs. plate size illustrating the grade of hierarchy of the <u>small plates</u> for the past 50 Ma. C: The same for the <u>large plates</u> for the past 140 Ma. Vertical fits (~110 Ma) correspond to maps with uniform plate tessellation (low $\alpha_{LP}$, weak hierarchy), while less vertical fits (~60 Ma) correspond to maps with contrasting colours (high $\alpha_{LP}$, strong hierarchy).

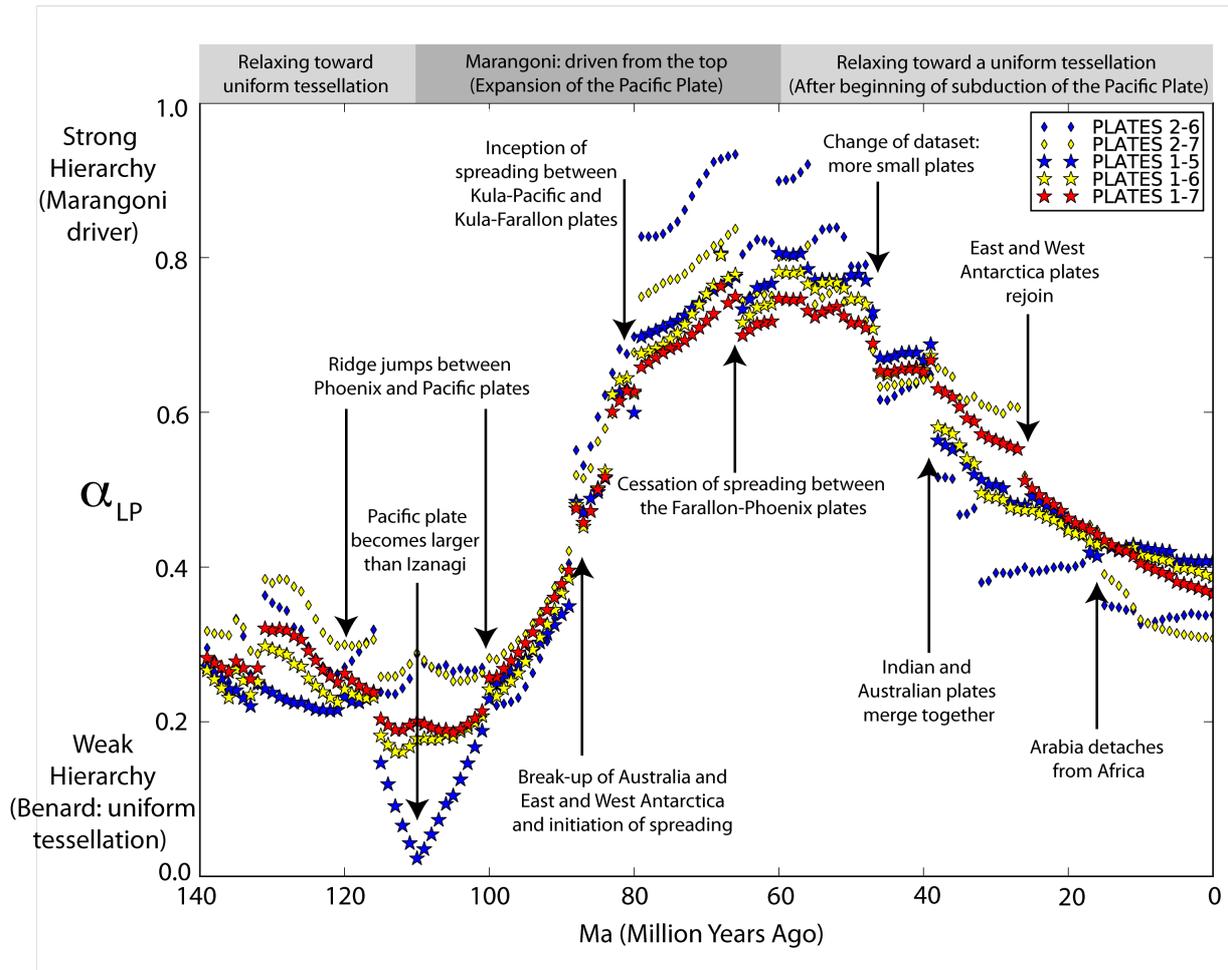

*Figure 3: $\alpha_{LP}$ every one million years. Five values are obtained by different overlapping subsets, as in Figure 1. The estimates of $\alpha_{LP}$ agree, in particular for the past 60 Ma, diverging only close to the extreme, when the system reorganises. Most jumps of $\alpha_{LP}$ are associated with fundamental geological events and accommodate the rapid growth and slow relaxation phases.*

*Supporting online Material for:*

Hierarchical self-organization of tectonic plates

*Gabriele Morra, Maria Seton and R. Dietmar Müller*

**Small Plates**

The Figures A1 and A4 show the same plots of figure 2b and 2c, but with the names of the plates and the indication in the legend of which fits are displayed. The Figure A2 is the equivalent of Figure 3 of the manuscript for the large plates. The Figures A3 and A5 display the residuals of the log-log fits (sum of the squared errors) for each fit.

The estimation of $\alpha_{SP}$ is based on six plate-size subsets for each 60 successive plate configurations: 9–18, 9–19, 9–20, 10–19, 10–20, 10–21. The 360 resulting fits are displayed in Figure A2. A detailed inspection of the names in Figure A1 shows how the small plates order, differently from the large plates structure, is extremely variable. Such a great variability in plate identification and number of small plates causes the intense variation in time illustrated in figure A2. This intense reorganisation, however, is characterised by a log-log linear fit constantly between 3 and 5 for the past 50 Ma, becoming steeper 60 Ma, when the dataset is more compact.

The residuals of the fits for the small plates are shown in Figure A3 and portray a relatively constant quality, always below 0.01. This value is the sum of the squared differences (res=sum{[y-f(x)]$^2$}); considering an average value around 1 in the y axis of the log-log plot, 0.01 corresponds to an average distance from the linear fit of about 3% (~100*sqrt[0.01/10]) for each of the 10 plates. Between 40 and 25 Ma, a minimum in of the residual is found, which corresponds to closer to 4 fits for the exponent $\alpha_{SP}$ (Figure A2), further confirming the fragmentation interpretation.

**Large Plates**

The fit of the largest plates is plotted in figure A4, with the plate names labelled. The five subsets are 1–5, 1–6, 1–7, 2–6, 2–7. The critical period at 120–100 Ma is characterised by the replacement of the dominant plate of the system, when the Pacific Plate becomes larger than the Izanagi Plate. This replacement starts the tectonic cycle. Between 40 and 25 Ma (see the 30 Ma plot) a less defined fit is due to the anomalous Indo-Australian plate, formed by joining Indian and Australian plates. This "anomaly" shows how the power law fit is an excellent identifier for anomalies in plate reconstructions.

The plot of the residuals for the largest plates reinforces the result and holds further indications. The top panel shows how the log-log fit was always excellent, with an anomalous peak characterised by high residuals only during the 100 Ma plate reorganisation, indicated by the swap between the Izanagi and Pacific Plates. The *zoom* in the bottom panel illustrates how the residuals are lower for present data, slowly increasing in past; this agrees with the consideration that the older is the plate reconstruction, the worse is the location of the plate boundary, confirming the importance of the power law fit as a predictor of plate reconstructions quality measure. This is confirmed as well by the more chaotic residuals for older reconstructions.

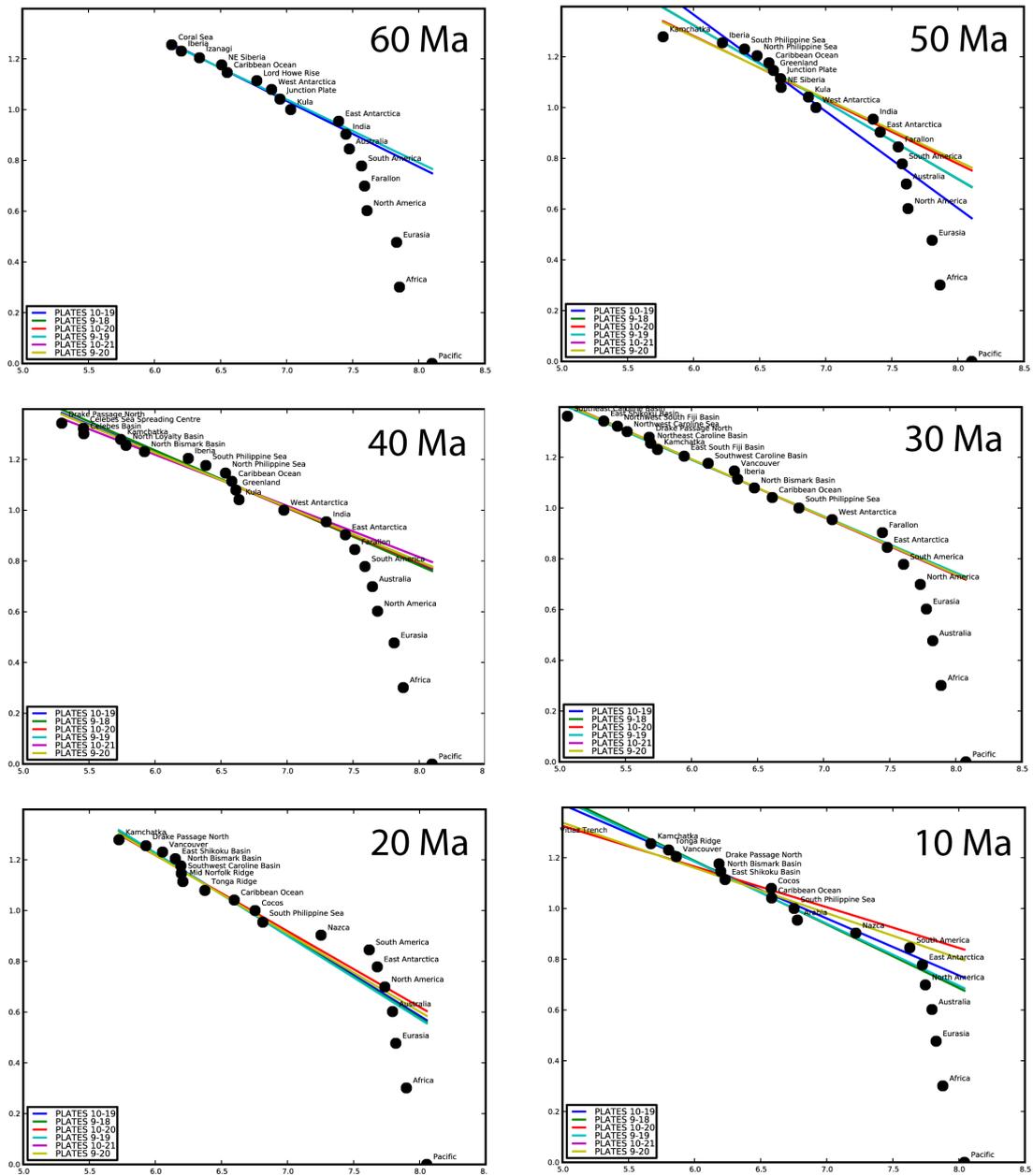

*Figure A1 – Detailed fit of the small plates from 60 Ma to 0 Ma, every 10 Myrs. The names of the plates are written besides every point. Six fits, one for each subset, as illustrated in the legend, are displayed.*

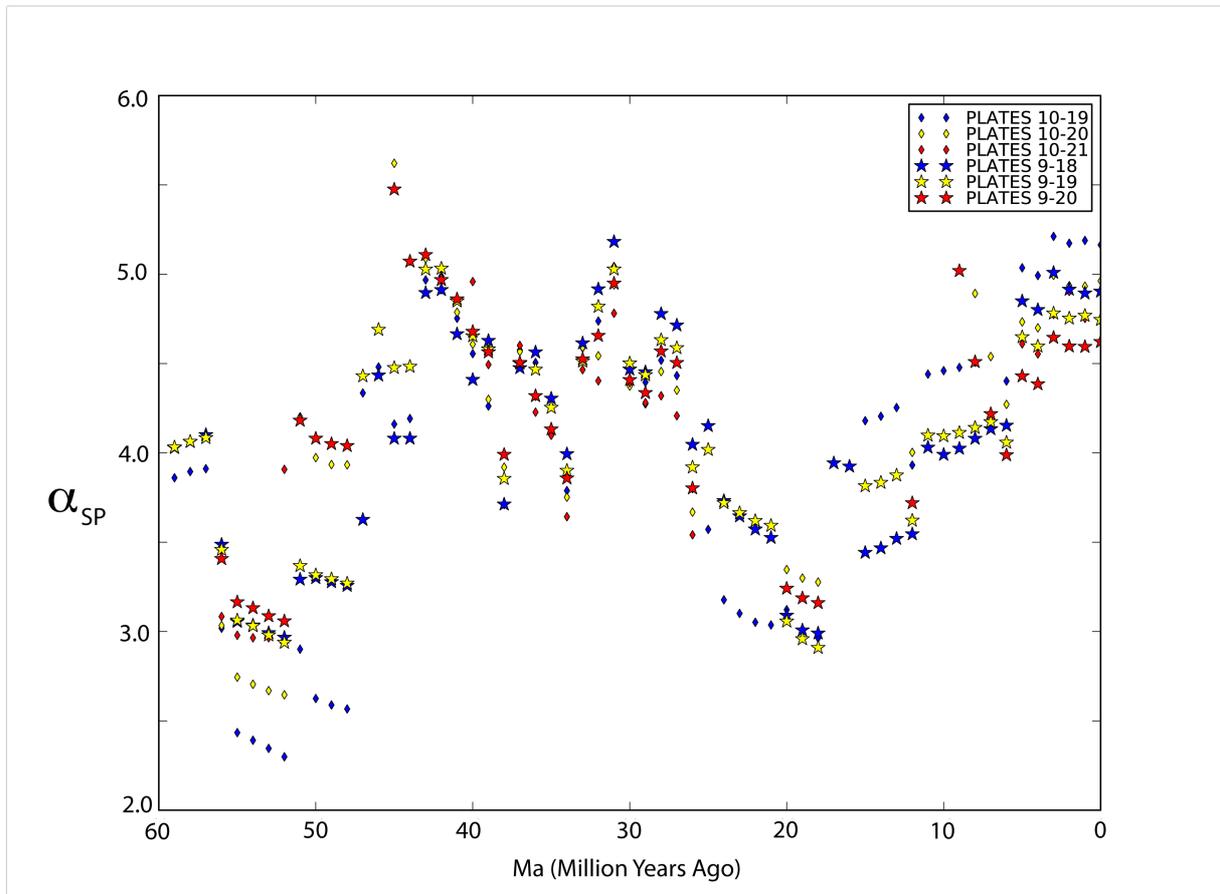

*Figure A2 – Time evolution of the fit for the small plates. There is not a pattern and the fits do not coincide for different subsets such as for the large plates. All fits vary between 3 and 5 in the past 50 Ma, oscillating around the value of 4 (fragmentation), and below three between 60 and 50, when the propagation of the errors of plate reconstruction has a greater effect.*

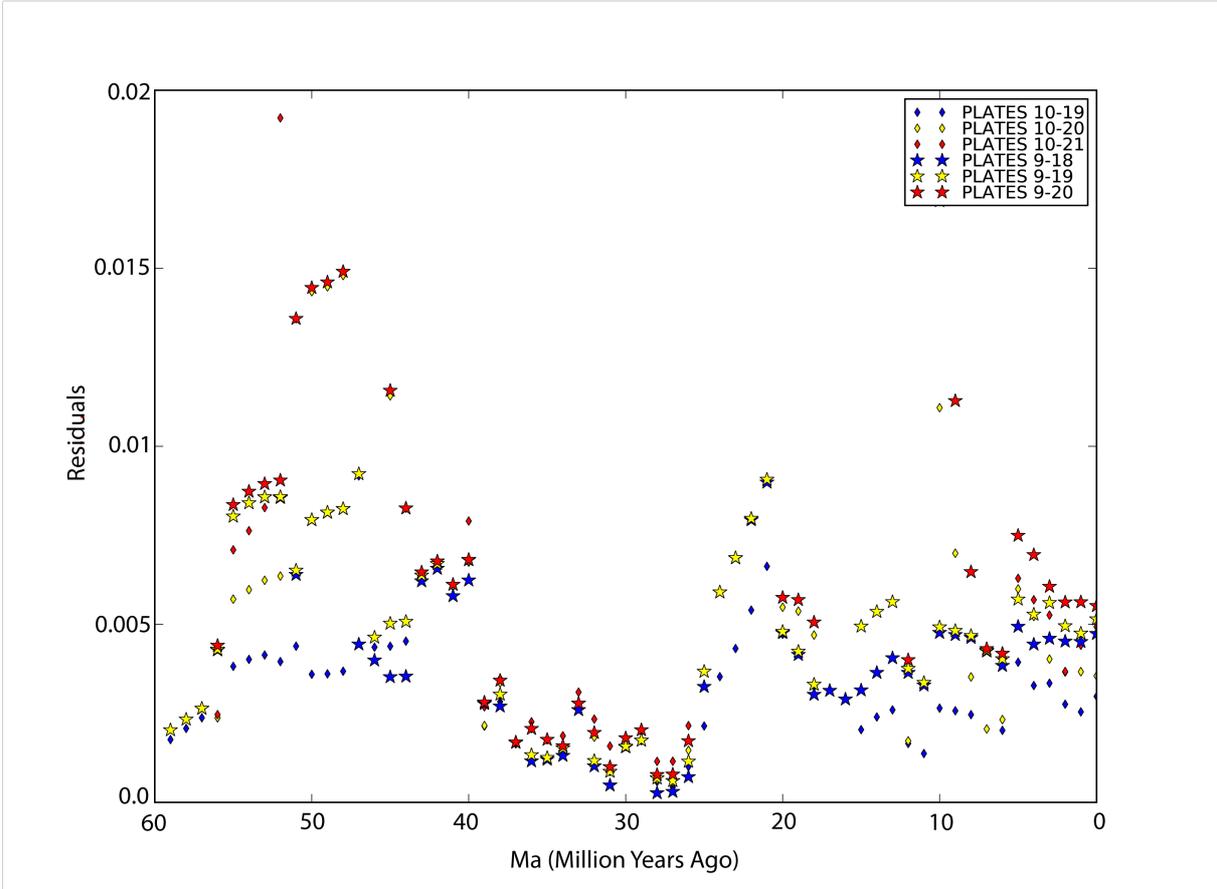

*Figure A3 Time evolution of the residues for the six fits of the small plate sizes displayed in figure A1 and A2. The residuals are mostly below 0.01, confirming the very good agreement with the data.*

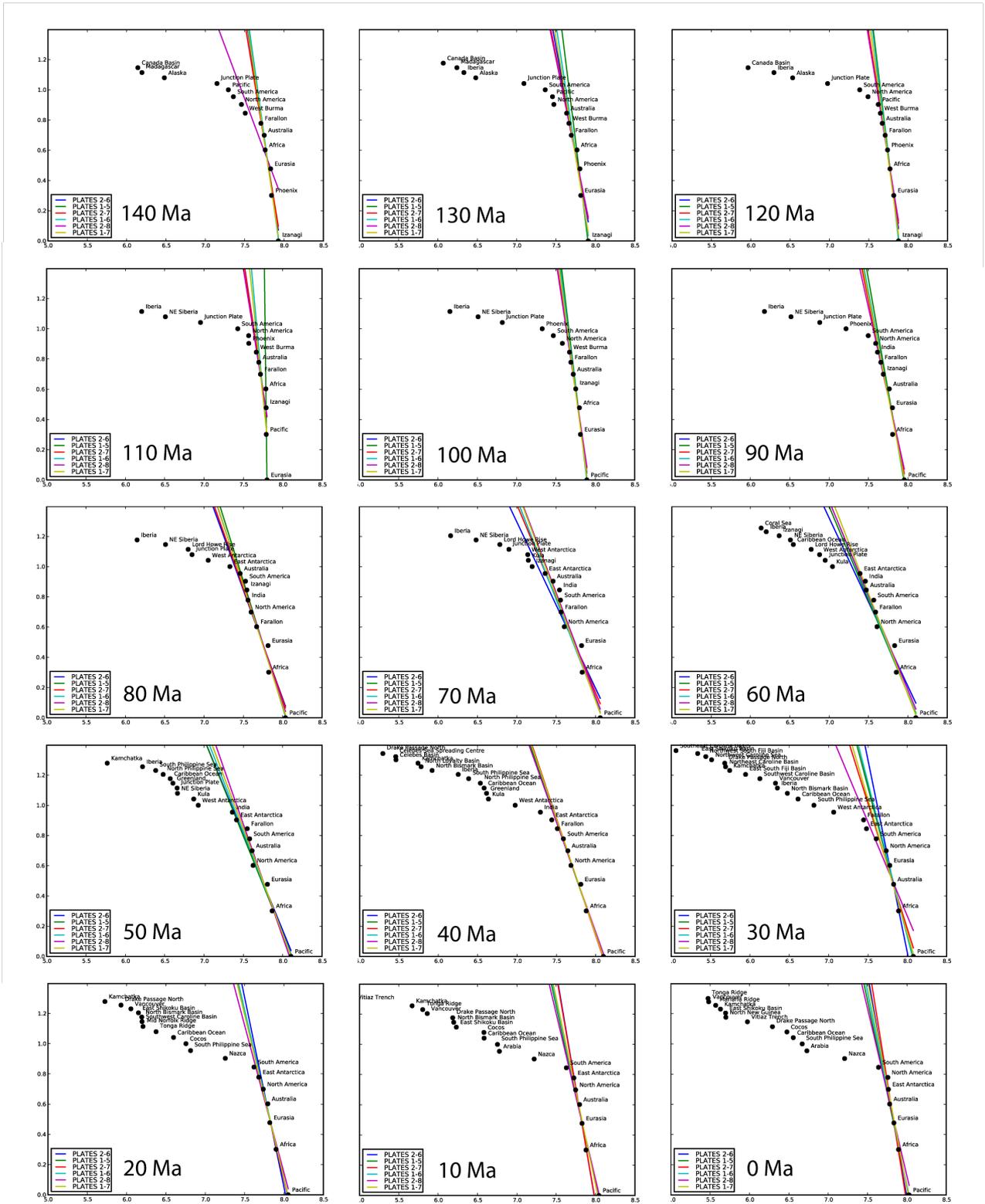

*Figure A4 – Fit of the largest plates from 140 to 0 Ma, every 10 Myrs, with the name of the plates.*

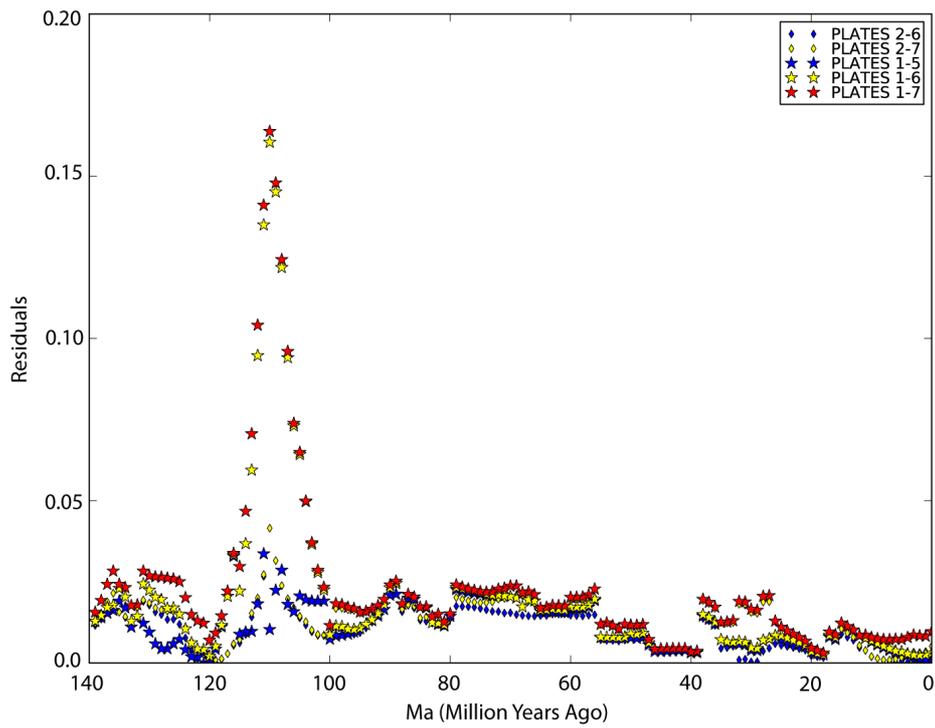
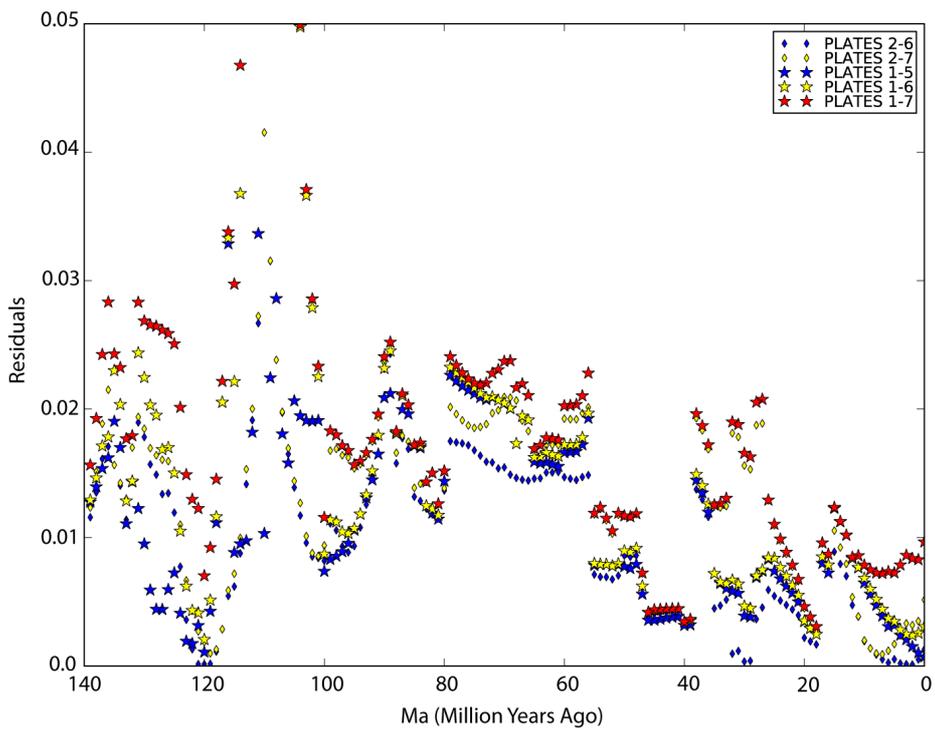

*Figure A5. Residuals for the large plate fits. In the top panel it is shown the values up to 0.2, while in the bottom panel the y-axis maximum values is reduced down to 0.05 with the aim to zoom on the evolution of the small residuals.*